# Deep-learning surrogate crop modelling for scalable seasonal-to-climate crop-risk assessment

O. Vlachopoulos[1], A. Ceglar[2], J. Luterbacher[1,3], N. Luther[1], A. Toreti[2], E. Xoplaki[4]

1 Department of Geography, Climatology, Climate Dynamics and Climate Change, Justus Liebig University Giessen, Giessen, Germany

2 European Commission, Joint Research Centre, Ispra, Italy

3 Liebig Centre for Agroecology and Climate Impact Research, Justus Liebig University of Giessen, Germany

4 CMCC Foundation - Euro-Mediterranean Center on Climate Change, Italy

# Abstract

Anticipating climate-related crop stress requires crop-risk information that is spatially explicit, probabilistic and fast enough for large seasonal forecast and climate-scenario ensembles. We present the Surrogate Engine for Crop Simulations Framework (SECSF), a deep-learning framework that emulates the process-based ECroPS model using only daily minimum and maximum temperature and precipitation. Trained on ERA5-forced ECroPS simulations for grain maize and spring barley, SECSF closely reproduces daily crop-growth dynamics and harvest timing while reducing computational cost by around four orders of magnitude, enabling ensemble-scale inference suitable for research and operational pipelines such as agricultural early warning and adaptation planning under uncertainty across seasonal-to-climate timescales. When forced with seasonal forecast data, SECSF captures spatially coherent crop-risk patterns across Europe in the high-impact year 2022 and is consistent with independent monitoring, supporting its use for probabilistic Areas of Concern. Under CMIP6 scenarios, SECSF identifies the Mediterranean basin as a hotspot of maize-risk signals through mid-century, with more mixed signals in central and northern Europe.

# Main

Reliable and timely assessment of climate-related crop stress at spatial and temporal scales relevant for preparedness, adaptation and early warning is increasingly important for resilient food production systems[1]. Process-based crop models are central to this task because they translate weather and environmental conditions into biologically meaningful crop responses. Models such as the Engine for Crop Parallelizable Simulations (ECroPS), World Food Studies (WOFOST) and the Agricultural Production Systems sIMulator (APSIM) are therefore widely used to assess crop performance, climate vulnerability and adaptation options under variable and changing environmental conditions [2–9]. In particular, they enable analysis of how crop varieties, cropping systems and agricultural practices may respond to changes in temperature, precipitation and the frequency and intensity of extreme events [10–13].

Artificial Intelligence (AI) is increasingly used to complement process-based crop modelling, particularly where input data, computational speed and scalability are limiting factors. In this context, surrogate models can emulate computationally intensive simulations while preserving key response patterns, opening the possibility of large ensemble applications, rapid scenario and sensitivity testing and operational deployment [14–18]. Here, we address this scalability barrier by developing an AI-based surrogate framework that translates process-based crop simulation into probabilistic crop-risk assessment at continental scale.

Recent crop growth emulation studies have shown that machine learning can emulate crop-growth outcomes and improve forecasting skill, including in hybrid process-based and AI frameworks [18–20]. However, many existing approaches focus primarily on final crop outcomes, rely on relatively input-intensive predictor sets, or do not explicitly reproduce the full daily evolution of crop development. This limits their ability to represent weather-sensitive crop responses between critical phenological phases, where cumulative and lagged effects of temperature and water stress can be important. In addition, the literature has not yet clearly established a streamlined surrogate framework that is simultaneously low input, process-informed, and directly usable for both seasonal early warning and climate-scale risk assessment. In this study, that capability is provided by a surrogate that emulates daily crop-growth dynamics using only daily minimum temperature, daily maximum temperature and daily precipitation. Recent developments such as CROMES further highlight the growing role of crop-model emulation in climate impact assessment, while also underscoring the need to define clearly what capability a given emulator provides [21]. The contribution of this study is to demonstrate a low-input, process-informed surrogate that reproduces daily crop-growth dynamics and can be deployed consistently across seasonal forecast and climate-scenario workflows. Here we present the Surrogate Engine for Crop Simulations Framework (SECSF), a set of deep-learning emulators trained on ECroPS simulations for grain maize and spring barley across Europe. Grain maize and spring barley are two major European crops with importance for feed, food, bioenergy and malting systems, making them relevant test cases for climate-related crop-risk assessment [22,23]. Their contrasting phenology and management relevance also provide a useful test of whether a reduced-input surrogate can reproduce crop-growth dynamics across different European production contexts. We show that SECSF closely reproduces daily crop-growth dynamics and harvest timing while reducing computational cost by around four orders of magnitude. We then demonstrate that this efficiency gain enables two applications central to spatially explicit crop-risk assessment: probabilistic Areas of Concern (AoC) based on seasonal forecasts, and continental-scale crop-risk assessment under Coupled Model Intercomparison Project Phase 6 (CMIP6) climate scenarios. By linking process-informed crop-growth emulation to ensemble climate information, the framework addresses a key food-system challenge: producing timely, spatially explicit and uncertainty-aware crop-risk evidence before impacts are fully realised.

# Results

We first assess whether SECSF reproduces crop-growth dynamics and harvest timing with sufficient fidelity to emulate the mechanistic ECroPS model. We then test whether the trained surrogate can be used with seasonal forecasts and climate-scenario forcings to generate spatially coherent crop-risk information across Europe. The framework is developed for two major European crops, grain maize and spring barley, with grain maize used below as the primary showcase for seasonal and climate-risk applications. SECSF comprises crop-specific emulators for grain maize (SECS4M) and spring barley (SECS4SB).

SECSF is forced with European Centre for Medium-Range Weather Forecasts (ECMWF) ReAnalysis version 5 (ERA5), ECMWF Seasonal Forecast System 5.1 (SEAS5.1) and CMIP6 climate data, with the ERA5-forced ECroPS output serving as the benchmark reference for the surrogate evaluation [24,25]. SEAS5.1 and CMIP6 forcing data are bias adjusted against ERA5 using quantile delta mapping before being used in the surrogate simulations [26].

## Performance

SECSF is trained on ERA5-forced ECroPS simulations for 1993–2023, using attainable Total Weight of Storage Organs (TWSO, kg $ha^{-1}$) under water-limited conditions as the target variable. Predictions are generated at daily resolution and at the grid-cell level, consistent with the ECroPS simulation framework. We therefore distinguish emulator fidelity, defined as agreement with ERA5-forced ECroPS trajectories, from application plausibility, defined as the ability of SECSF to generate spatially coherent and externally interpretable crop-risk patterns under seasonal forecast and climate-scenario forcing.

### Computational performance

SECSF is computationally lightweight enough for large-scale ensemble applications. On the Deutsches Klimarechenzentrum (DKRZ, German Climate Computing Centre) Levante system, SECSF requires approximately 0.008 seconds of CPU time to generate a one-year daily TWSO trajectory for one grid cell, compared with approximately 70 seconds for ECroPS under the same configuration. This corresponds to a speed-up of roughly four orders of magnitude and is central to the feasibility of seasonal ensemble forecasting and climate-scenario crop-risk applications.

### Goodness of fit of SECSF against ECroPS

SECSF is evaluated in two steps. First, the emulator is trained and tested against ERA5-forced ECroPS simulations, using attainable TWSO as the dependent variable. Second, after training, the surrogate is forced with bias-adjusted SEAS5.1 and CMIP6 weather data and compared with the corresponding ERA5-forced surrogate reference, upscaled to match the SEAS5.1 or CMIP6 spatial resolution. Direct ECroPS simulations with SEAS5.1 and CMIP6 were not conducted because their grid definitions and spatial resolutions are incompatible with the EC-JRC crop, soil and agro-management inputs used here. These necessary static inputs are provided at approximately 0.25°, closely matching the ERA5 spatial grid and thus allowing the use of ERA5 input weather variables in order to generate simulations. ECroPS is spatial resolution agnostic, and all input grids should be similar if not identical. As a result, ECroPS cannot be run with SEAS5.1 or CMIP6 data using their native grid along the static inputs from EC-JRC. These alternative forcing streams are applied only through the surrogate which simulates using only three weather inputs of identical grids. The evaluation therefore separates two distinct questions: whether SECSF faithfully reproduces ECroPS behaviour under ERA5 forcing, and whether alternative climate forcings generate spatially coherent and physically plausible crop-risk patterns when passed through the trained surrogate. This evaluation is therefore not a full observational TWSO validation across all applications, but an assessment of emulator fidelity, forcing-dependent coherence and crop-risk plausibility.

In the first evaluation step, Fréchet and Hausdorff distances quantify geometric similarity between predicted and ECroPS-reference time series (Figure A1). Figure 1 complements these metrics by showing representative daily crop-growth trajectories for grain maize and spring barley. In both crops, SECSF closely follows the temporal evolution of TWSO and the timing of maturity, indicating that the surrogate can reproduce growth trajectories rather than only final TWSO levels.

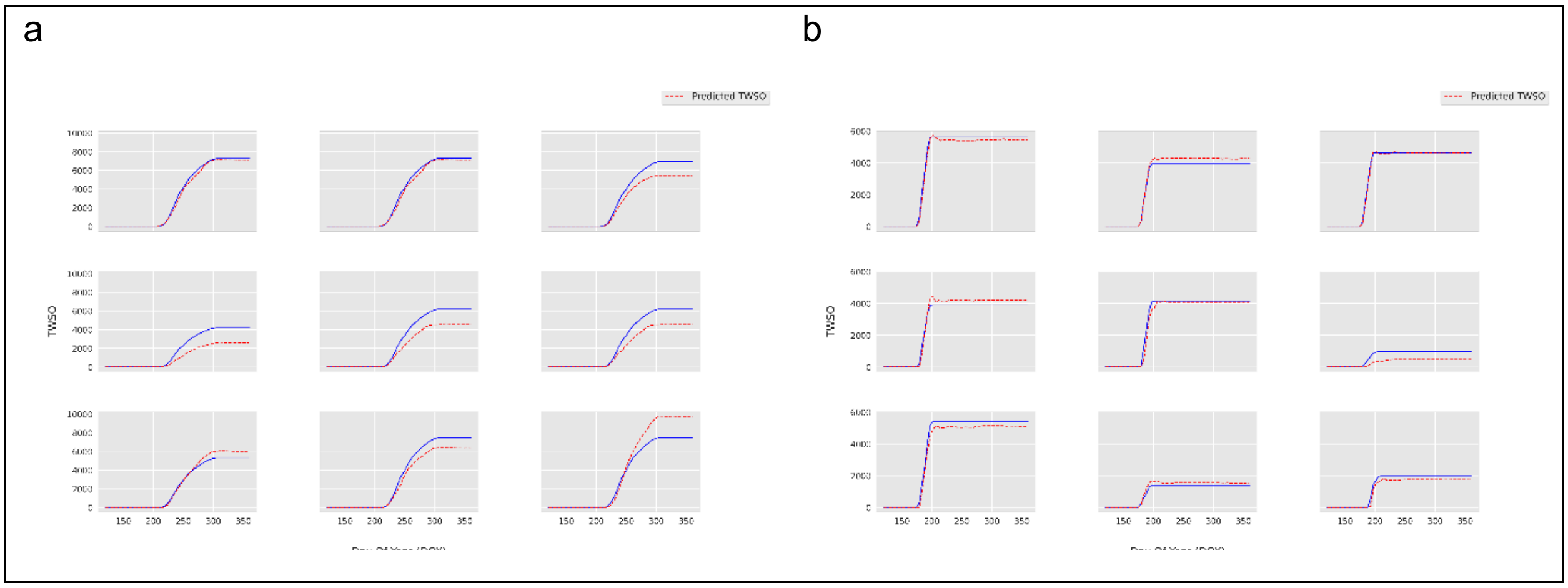

*Figure 1: Representative daily crop-growth trajectories for grain maize (a) and spring barley (b), showing ECroPS-reference and SECSF-predicted attainable Total Weight of Storage Organs (TWSO, kg ha $^{-1}$) at selected grid cells.*

## Seasonal forecast and CMIP6 forcing of SECS4M

Grain maize is used as the primary showcase crop for evaluating SECSF under seasonal forecast and climate-scenario forcing. In the following, SECS4M outputs driven by bias-adjusted SEAS5.1 and CMIP6 data are compared with the corresponding ERA5-forced surrogate reference to assess whether the trained emulator produces coherent spatial and temporal crop-response patterns across different climate inputs.

a b

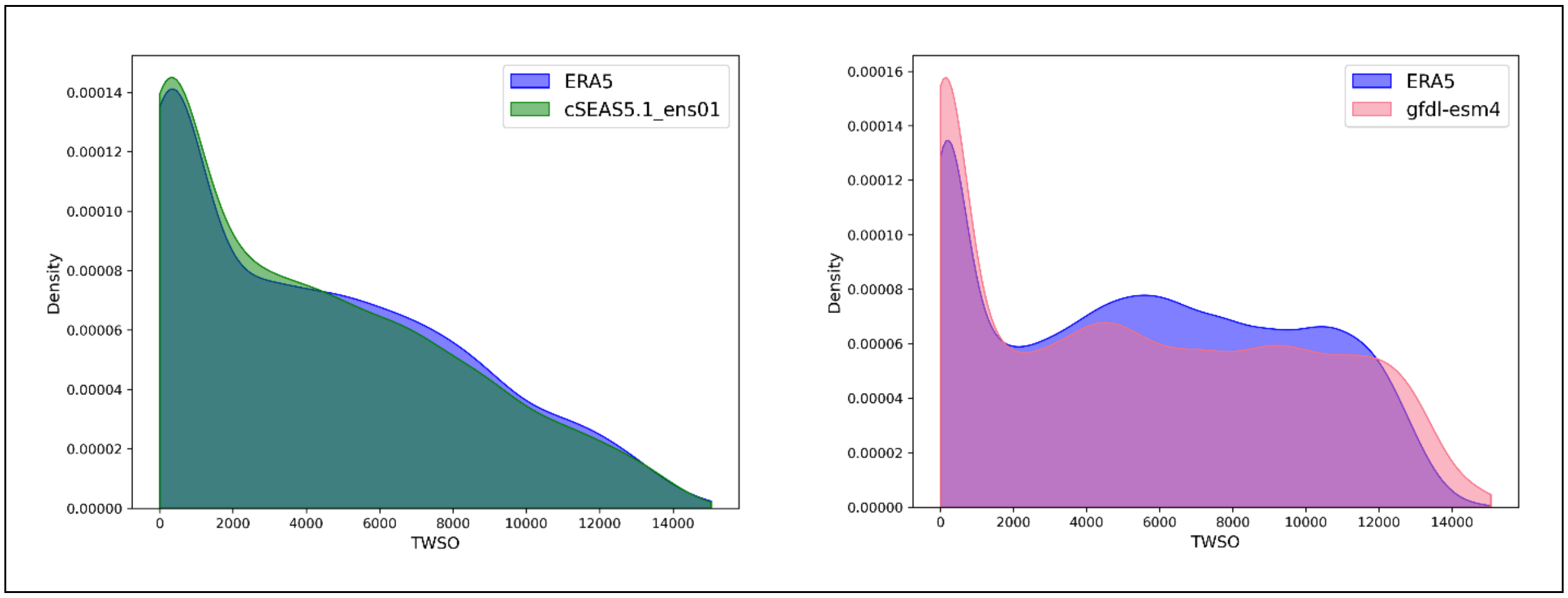

*Figure 2: Probability density functions of SECS4M-simulated attainable Total Weight of Storage Organs (TWSO, kg ha⁻¹) under ERA5 forcing and under alternative climate forcing for 1993–2014. Panel (a) compares ERA5-forced SECS4M output with bias-adjusted SEAS5.1 ensemble member 01 (cSEAS5.1_ens01). Panel (b) compares ERA5-forced SECS4M output with GFDL-ESM4 historical realization r1i1p1f1.*

## SEAS5.1

For SEAS5.1, SECS4M simulations are produced for both the reforecast period (1993–2016; 25 ensemble members) and the forecast period (2017–2023; 51 ensemble members). We focus on the June 2022 forecast because 2022 was a year of severe grain-maize stress and production loss in Europe [27]. To emulate an operational full-season application, SECS4M is forced with ERA5 from January to May and with bias-adjusted SEAS5.1 from June to December, hereafter termed cSEAS5.1. This setup is relevant because the summer months correspond to lead times 0 to 2, when forecast skill is expected to be highest and weather conditions are particularly consequential for maize development.

Figure 2a shows that the ERA5- and cSEAS5.1-forced TWSO distributions retain a common overall structure, with modest differences in the representation of intermediate and extreme values. This indicates that the seasonal forecast forcing preserves the broad distributional behaviour of the ERA5-forced surrogate reference while introducing forcing-dependent differences that require spatial assessment.

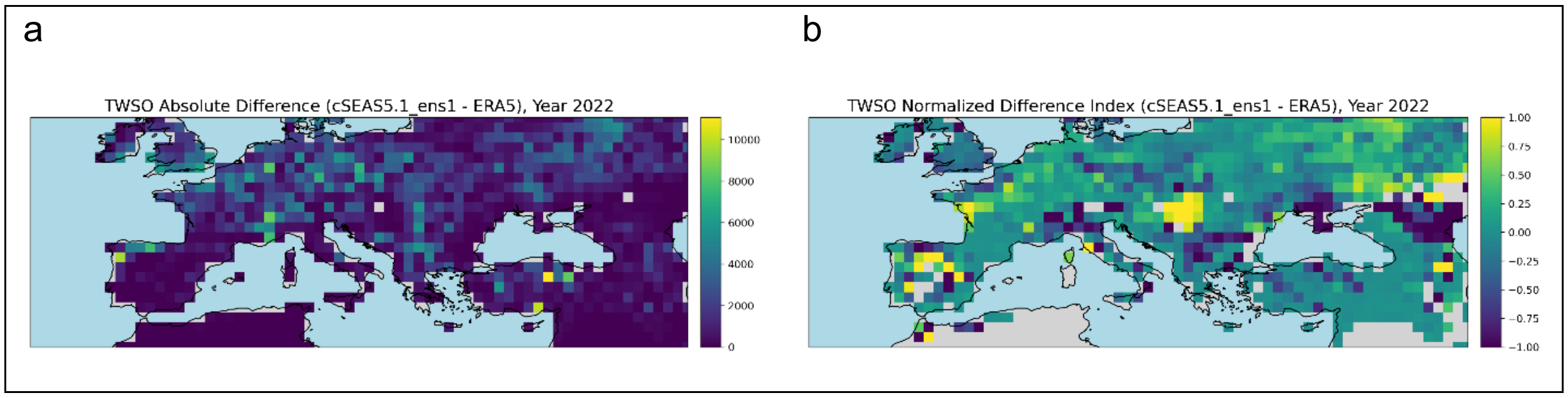

*Figure 3: Spatial comparison of ERA5- and cSEAS5.1_ens01-forced SECS4M TWSO for 2022, showing (a) the absolute difference between cSEAS5.1 and ERA5 forcing (cSEAS5.1 minus ERA5; kg ha⁻¹) and (b) the normalized difference index.*

Figure 3 shows that the seasonal-forcing application preserves the broad spatial structure of simulated 2022 maize outcomes while revealing coherent regional differences between

cSEAS5.1- and ERA5-forced TWSO. Absolute deviations are generally modest, but spatially organized hotspots indicate localized sensitivity to forcing differences.

The sign of the difference is also spatially structured, with cSEAS5.1 tending to produce relatively higher TWSO in cooler and wetter parts of Europe and relatively lower TWSO in warmer and drier regions. This pattern suggests that SECSF can retain geographically coherent forcing-dependent crop-response signals, rather than only domain-mean anomalies. Grid cells outside the valid crop-simulation domain are masked throughout.

### CMIP6

For the CMIP6 evaluation, SECS4M is forced with the historical simulations listed in Table A1 to generate TWSO outputs for 1993–2014, matching the evaluation window used for SEAS5.1. Representative diagnostics are shown using GFDL-ESM4 as an illustrative CMIP6 example. Returning to the distributional comparison in Figure 2b, both datasets share a strongly right-skewed structure, but the GFDL-ESM4-forced simulation assigns greater probability to very low and very high TWSO values, whereas the ERA5-forced simulation is more concentrated in the mid-range. This indicates that the CMIP6 forcing preserves the broad response shape while producing a wider spread of simulated crop outcomes. Consistent with this distributional contrast, the CMIP6-forced evaluation shows that SECS4M retains the broad climatological response structure of the ERA5-forced surrogate reference while also producing spatially heterogeneous forcing-dependent differences (Figure 4). Most of western and central Europe exhibit relatively small long-term deviations, whereas more pronounced mixed-sign differences occur in localized eastern regions. Together, Figure 2b and Figure 4 indicate that the trained surrogate preserves the large-scale response structure while remaining sensitive to regionally coherent differences in the driving climate data.

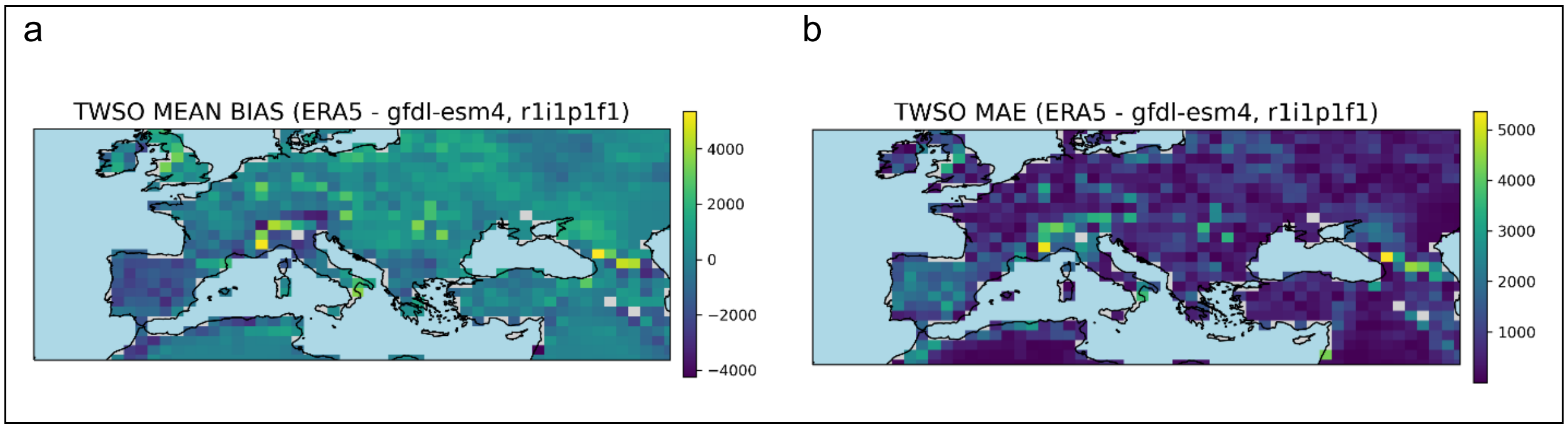

*Figure 4: Spatial comparison of ERA5- and GFDL-ESM4 r1i1p1f1-forced SECS4M TWSO for 1993–2014, showing (a) mean bias calculated as ERA5 minus GFDL-ESM4 and (b) mean absolute error (kg ha $^{-1}$).*

## Probabilistic Areas of Concern

### cSEAS5.1

For cSEAS5.1-forced SECS4M outputs, AoC detection follows a tercile-based probabilistic protocol. Figure 5 shows, for 2022, the most probable TWSO anomaly category at each grid cell based on terciles derived from the 1993–2016 SEAS5.1 reforecast reference period. The 2022 AoC map identifies a broad belt of maize concern extending from western and central Europe into south-eastern Europe, while the Atlantic northwest and parts of northern Europe remain mostly normal to above normal (Figure 5). This spatial pattern is consistent with contemporaneous monitoring from JRC MARS, Copernicus ESOTC, and the European Commission, which reported strong maize stress and production losses across France, Italy, and central to south-eastern Europe in 2022 [27–31]. This agreement supports the use of the surrogate framework to translate seasonal climate information into probabilistic crop-risk products for high-impact years, while remaining conditional on the ECroPS-based training framework and the tercile-based AoC definition.

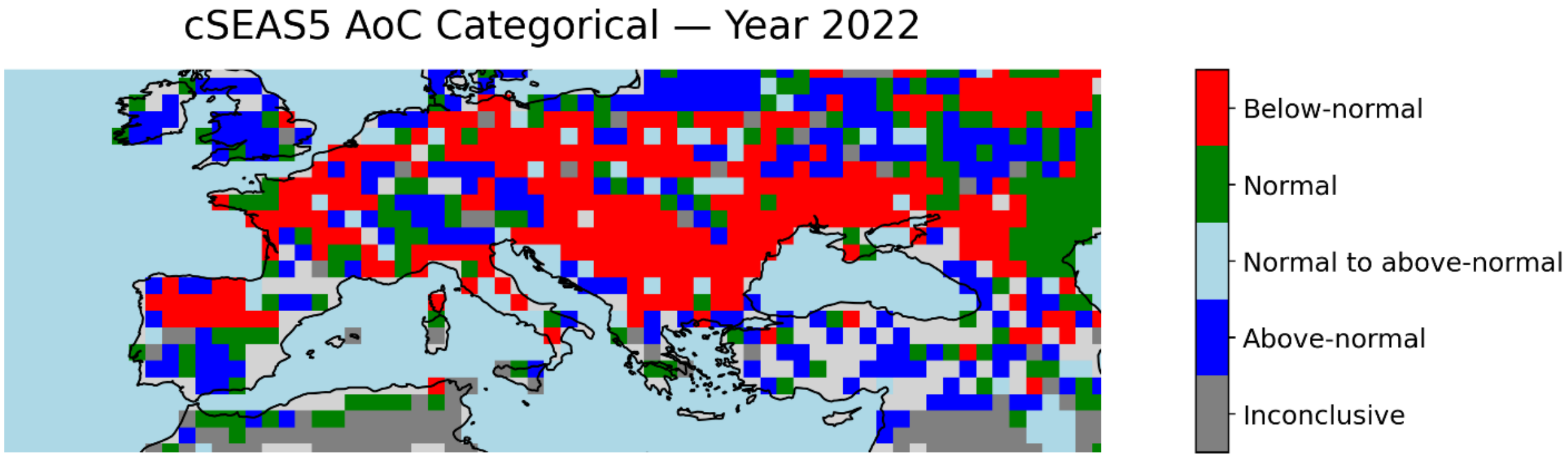

*Figure 5: Probabilistic Areas of Concern for grain maize in 2022, derived from 51 cSEAS5.1 ensemble members. Colours indicate the most probable attainable Total Weight of Storage Organs (TWSO) anomaly category at each grid cell, classified relative to terciles derived from the 1993–2016 SEAS5.1 reforecast reference period. The AoC classes used are below-normal, normal, normal-to-above-normal, above-normal and inconclusive.*

## CMIP6

For CMIP6, the AoC analysis is conducted using multiple realizations under SSP3-7.0 and SSP5-8.5 (Table A2). Each future simulation is paired with the corresponding historical realization, which is used as the reference baseline for relative TWSO anomalies. The analysis assumes unchanged production systems and does not include explicit adaptation, technological change or altered management. It is therefore restricted to the mid-century horizon, beyond which fixed production-system assumptions would become increasingly unrealistic.

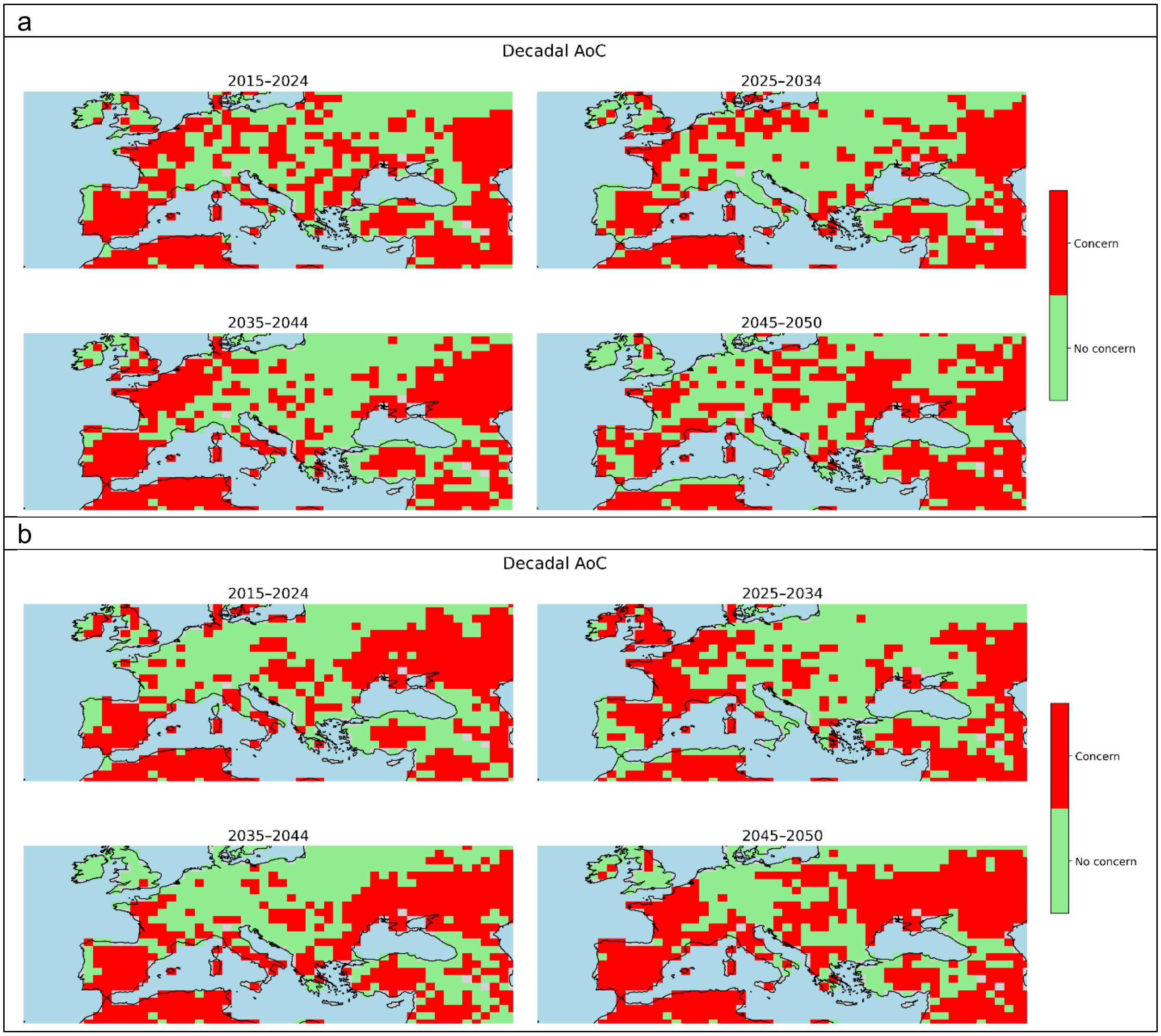

*Figure 6: Decadal-average Areas of Concern for grain maize derived from GFDL-ESM4-forced SECS4M TWSO under (a) SSP3-7.0 and (b) SSP5-8.5. Colours indicate the dominant AoC category for each decade, based on TWSO anomalies relative to the corresponding historical GFDL-ESM4 baseline.*

Across both SSP3-7.0 and SSP5-8.5, the AoC analysis for the chosen example of GFDL-ESM4 indicates a persistent concentration of maize risk in the Mediterranean basin and adjacent Black Sea regions through mid-century (Figure 6). The spatial footprint expands inland most clearly in the 2035–2050 period, while northern and parts of central Europe retain more mixed or intermittently lower concern. The higher-emissions pathway produces a more persistent and contiguous AoC signal in this realization, especially in the central-eastern Mediterranean and Black Sea rim. Taken together, these illustrative results identify the Mediterranean basin as a recurrent hotspot of future maize-risk signals under unchanged production-system assumptions.

# Discussion

This study shows that a low-input deep-learning surrogate can reproduce the daily crop-growth dynamics of a mechanistic crop model while reducing computational cost by around four orders of magnitude. This combination of fidelity and speed is important because it makes ensemble-scale crop-risk assessment feasible across both seasonal forecasting and climate-scenario applications. SECSF therefore provides a computational bridge between mechanistic crop simulation and spatially explicit agricultural risk products that can be generated at ensemble scale. The main contribution of SECSF lies in combining four properties that are rarely available together: consistency with process-based crop-growth dynamics, a very low meteorological input requirement, computational efficiency compatible with large ensembles, and applicability across both seasonal forecasting and climate-scenario contexts. In the present study, this combination enables probabilistic AoC mapping for the high-impact year 2022 and identifies a persistent mid-century concentration of maize risk in the Mediterranean basin under fixed production-system assumptions. From a decision-support perspective, such information can help identify stress-prone regions, seasonal windows of concern and candidate areas where adaptation options such as altered sowing dates, stress-tolerant cultivars or targeted resource allocation may warrant closer assessment, consistent with the growing emphasis on geographically explicit crop-risk assessment in agricultural risk management [32]. More broadly, the framework shows that surrogate modelling can move beyond retrospective TWSO approximation toward operationally relevant agricultural early warning and climate-risk assessment. For food-system risk management, this is important because rapidly generated crop-stress information can support earlier identification of vulnerable production regions and more targeted preparedness actions contributing to broader efforts to strengthen food-system monitoring and resilience[33]. Finally, this work has been implemented as an operational toolset in the Free Evaluation System Framework (Freva), an open-source platform for data discovery and analysis developed jointly by Freie Universität Berlin and DKRZ.

Several limitations are important for interpretation. First, SECSF inherits the structural assumptions, parameter choices and domain coverage of the ECroPS simulations on which it is trained. Its credibility therefore depends on the realism of the training archive, particularly for rare or underrepresented extremes. The present evaluation demonstrates emulator fidelity to ECroPS and the coherence of forcing-dependent crop-risk patterns, but it does not replace direct observational validation of yield impacts. Second, although SECSF reproduces ECroPS behaviour, it does not retain the full mechanistic transparency of the underlying process-based model. Third, the current implementation is region-specific because it is trained on EC-JRC valid cells and associated static soil, crop and management inputs, which limits immediate transferability beyond the present European domain. The future AoC patterns should therefore be interpreted as climate-driven risk signals under fixed production-system assumptions, not as projections of realised future yields under adaptation, technological change or altered management. Future work should extend the framework to additional major crops such as winter wheat, incorporate nutrient limitations and management settings, and explore coupling with real-time Earth observation and higher-resolution forecast information to strengthen independent evaluation and operational risk assessment. In summary, the significance of SECSF is not only that it accelerates crop simulation, but that it changes the scale at which process-informed crop-risk assessment can be deployed. By making seasonal and climate-timescale applications computationally tractable with minimal meteorological inputs, the framework opens a practical route toward more responsive crop-risk monitoring and more scalable agricultural climate services.

# Methods

## Model inputs and forcing data

ECroPS is used to simulate daily grain maize and spring barley growth over the European domain employed by the EC-JRC MARS Crop Yield Forecasting System for 1993–2023 [11,34]. The model configuration represents regional crop-parameter variability across Europe, including differences in variety prevalence. ECroPS can simulate potential, water-limited, and nutrient-limited production; here we focus on the water-limited setup as a practically relevant and computationally tractable configuration for surrogate development and evaluation.

ECroPS is forced with six daily meteorological variables: maximum temperature, minimum temperature, surface downwelling shortwave radiation, total precipitation, relative humidity and wind speed derived from vector wind components.

Soil characteristics are provided as gridded static EC-JRC inputs. The European soil database is organized into soil mapping units containing one or more soil typological units, from which properties such as rooting depth, available water capacity and infiltration capacity are derived. To avoid multiple simulations per grid cell and maintain a scalable setup, each grid cell is represented here by its dominant soil type.

Crop-specific and agro-management inputs are provided per grid cell by the EC-JRC, including sowing date, thermal-time requirements for key phenological phases, vernalization parameters and dry-matter partitioning coefficients. Crop and soil parameters from EC-JRC are stationary over the years, whereas ERA5 provides the time-varying meteorological forcing for 1993–2023. Each grid cell is simulated independently from sowing to the end of the calendar year, producing daily outputs including attainable Total Weight of Storage Organs (TWSO).

The surrogate reduces the meteorological and static input feature space to three variables: daily minimum temperature, daily maximum temperature and daily total precipitation. These variables are selected because they are directly linked to temperature and water-related crop stress while enabling a simpler and more transferable setup than the six-variable ECroPS forcing configuration.

## AI pipeline, architecture and feature engineering

SECSF uses a nested Long Short-Term Memory (LSTM) architecture to model nonlinear crop responses to sequential weather inputs. This choice is motivated by the need to preserve both short-term weather variability and longer-term temporal dependencies in crop development. The nested structure allows inner sequence blocks to capture short-term patterns, while outer recurrent layers retain broader temporal evolution signals across the season. The hierarchical design enables the network to capture temporal abstraction levels, effectively modeling detailed short-term fluctuations and broader long-term trends simultaneously [35,36].

For training and evaluation, the ERA5-forced ECroPS dataset is partitioned temporally, using all samples from 1993 to 2016 for training and all samples from 2017 to 2023 for testing. Each sample represents a specific location with daily weather observations over multiple years and the corresponding ECroPS-simulated TWSO target. The predictor feature space comprises daily maximum and minimum temperature, total precipitation, and engineered features including lagged weather inputs and day of year. Lagged weather features are included to capture persistence in weather regimes, delayed crop responses to cumulative temperature and precipitation effects, and reoccurring sequential patterns such as dry spells followed by rainfall. For each day, the model incorporates weather conditions from the previous one to five days.

Input sequences are organized into non-overlapping 6-day batches. This design provides sufficient local context to capture weather-driven crop responses while remaining computationally efficient. Within the nested architecture, an inner TimeDistributed LSTM learns short-term patterns from each batch, and an outer LSTM integrates these representations across the season to model longer-term crop development (Figure A2).

The output layer uses a TimeDistributed dense layer with ReLU activation to generate sequential non-negative TWSO predictions. A dropout rate of 0.3 is used to limit overfitting. Training employs the Adam optimizer, 128 units per LSTM layer, and a Huber loss with $\delta = 0.5$ [37,38]. All components are implemented in Python using the TensorFlow Keras framework.

SECSF performance on the testing dataset is evaluated using discrete Fréchet and Hausdorff distances [39,40]. The Fréchet distance captures trajectory similarity while preserving the temporal order of points, whereas the Hausdorff distance emphasizes maximum structural deviation. Together, they provide complementary information on the fidelity of predicted crop-growth trajectories.

Agreement between ERA5-forced and CMIP6- or cSEAS5.1-forced TWSO is further assessed using kernel-density estimates based on all available grid cells and time steps, allowing comparison of central tendency, spread and tail behaviour. Spatial forcing-dependent differences are evaluated using maps of mean bias and mean absolute error (MAE).

## Areas of Concern

SECSF output is further used to identify AoC, defined here as regions with elevated likelihood of adverse crop outcomes under unfavourable climate or weather conditions. A related concept is used by the EC-JRC to highlight areas at risk of substantial agricultural impacts [41,42].

We use a simplified AoC framework tailored to probabilistic crop-risk assessment. AoC are identified from SECSF outputs as TWSO anomalies relative to a reference period, with ensemble-based classification derived from terciles of the historical anomaly distribution. For cSEAS5.1, the reference is based on the full SEAS5.1 reforecast ensemble for 1993–2016. For CMIP6, future anomalies are computed relative to the corresponding historical realization for each model and scenario before applying the same tercile-based AoC classification. Forecast-period anomalies are computed against the same baseline and classified into tercile-based categories. Grid-cell probabilities are then derived from the fraction of ensemble members assigned to each category, and the category with the highest ensemble probability is used to generate the AoC maps shown in the Results. The decision rules are: (i) “inconclusive” if probabilities of below-normal and above-normal are equal; (ii) “above-normal” when above-normal probability exceeds the others; (iii) “normal to above-normal” when above-normal and normal probabilities are equal; (iv) “normal” if normal probability surpasses others; and (v) “below-normal” when below-normal probability is highest.

## Author Contributions

Conceptualization, O.V., N.L., A.C., A.T. and E.X.; Data curation, O.V.; Investigation, O.V., N.L. and E.X.; Methodology, O.V.; Software, O.V.; Writing – original draft, O.V.; Writing – review and editing, O.V., A.C., J.L., A.T. and E.X. All authors have read and approved the submitted version of the manuscript.

## Acknowledgments

The authors thank Stefan Niemeyer (EC-JRC) and Davide Fumagalli (EC-JRC) for their support in providing crop-related data for ECroPS. O.V., N.L. and E.X. acknowledge support from the Horizon 2020 project CLINT (number 101003876) and the German Federal Ministry of Research, Technology and Space (BMFTR) project ClimXtreme II.

# Extended Data

*Table A1: CMIP6 historical simulations used to force SECS4M for the 1993–2014 historical evaluation period*

| Model name | Historical realizations | Approximate nominal resolution in degrees (latitude, longitude) | Reference |
|---|---|---|---|
| CNRM-CM6-1-HR | 'r1i1p1f2' | 0.5°, 0.5° | [43] |
| EC-Earth3 | 'r11i1p1f1', 'r13i1p1f1', 'r15i1p1f1', 'r1i1p1f1', 'r3i1p1f1', 'r4i1p1f1', 'r6i1p1f1', 'r9i1p1f1' | 0.7°, 0.7° | [44] |
| GFDL-ESM4 | 'r1i1p1f1' | 1°, 1.25° | [45] |
| HadGEM3-GC31-MM | 'r1i1p1f3' | 0.56°, 0.8° | [46] |
| MPI-ESM1-2-HR | 'r10i1p1f1', 'r1i1p1f1', 'r2i1p1f1', 'r3i1p1f1', 'r4i1p1f1', 'r5i1p1f1', 'r6i1p1f1', 'r7i1p1f1', 'r8i1p1f1', 'r9i1p1f1' | 1°, 1° | [47] |
| NorESM2-MM | 'r1i1p1f1' | 1°, 1.25° | [48] |

*Table A2: CMIP6 future realizations used to force SECS4M for TWSO simulation and future Areas of Concern (AoC) generation*

| Model name | SSP3-7.0 | SSP5-8.5 |
|---|---|---|
| CNRM-CM6-1-HR | - | 'r1i1p1f2' |
| EC-Earth3 | 'r11i1p1f1', 'r13i1p1f1', 'r15i1p1f1', 'r1i1p1f1', 'r6i1p1f1', 'r9i1p1f1' | 'r11i1p1f1', 'r13i1p1f1', 'r15i1p1f1', 'r1i1p1f1', 'r3i1p1f1', 'r4i1p1f1', 'r6i1p1f1', 'r9i1p1f1' |
| GFDL-ESM4 | 'r1i1p1f1' | 'r1i1p1f1' |
| HadGEM3-GC31-MM | - | 'r1i1p1f3' |
| MPI-ESM1-2-HR | 'r10i1p1f1', 'r1i1p1f1', 'r2i1p1f1', 'r3i1p1f1', 'r4i1p1f1', 'r5i1p1f1', 'r6i1p1f1', 'r7i1p1f1', 'r8i1p1f1', 'r9i1p1f1' | 'r1i1p1f1', 'r2i1p1f1' |
| NorESM2-MM | 'r1i1p1f1' | 'r1i1p1f1' |

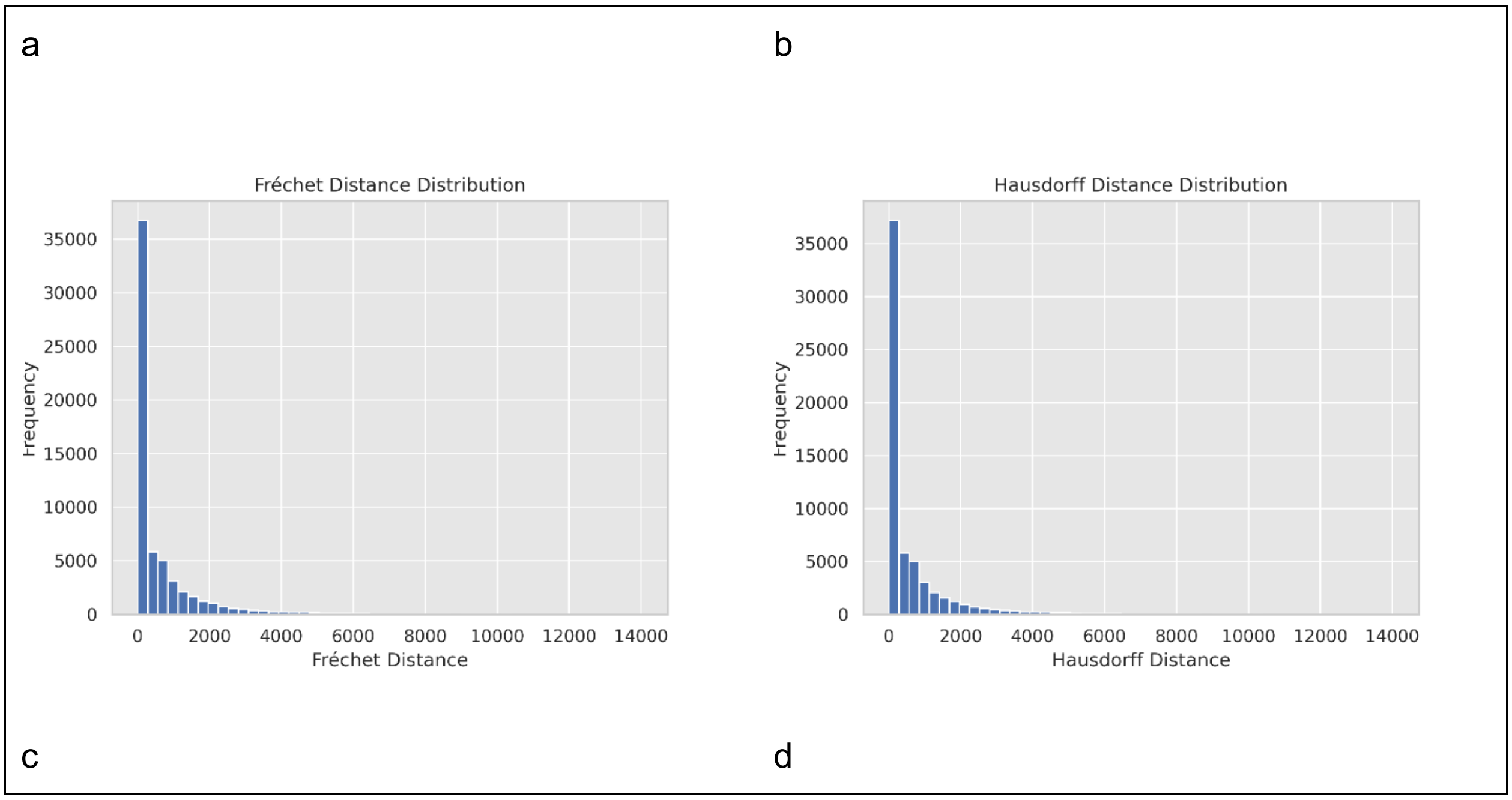

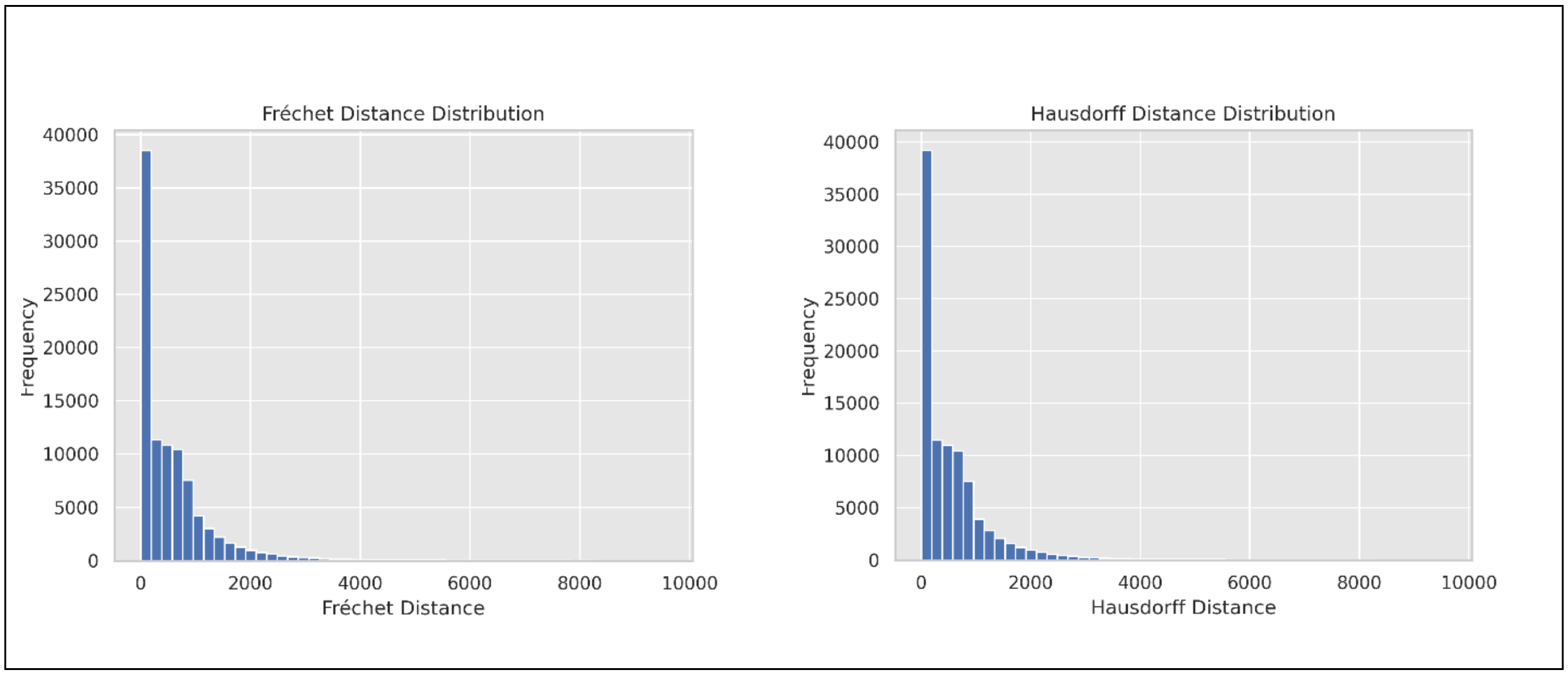

*Figure A1: Histograms of discrete Fréchet and Hausdorff distances for grain maize (a, b) and spring barley (c, d), summarizing agreement between SECSF-predicted and ECroPS-reference daily TWSO trajectories.*

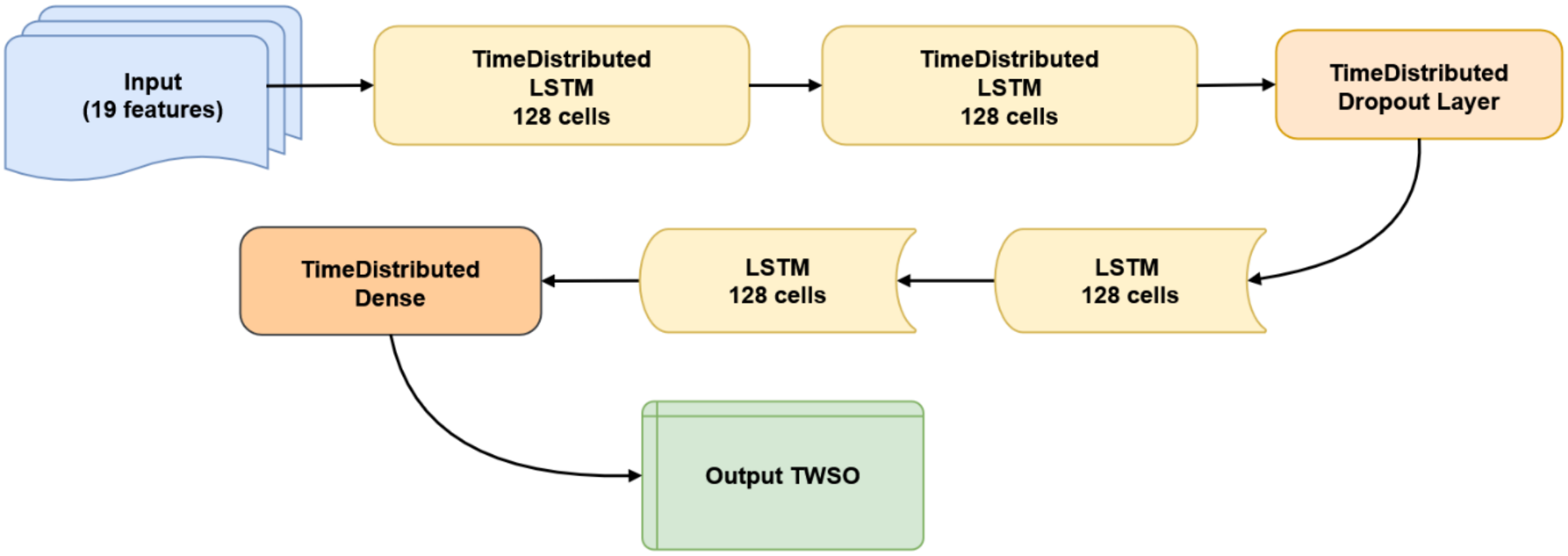

*Figure A2: Schematic representation of the nested LSTM architecture used in SECSF. Input weather sequences are organized into non-overlapping 6-day batches, processed by an inner TimeDistributed LSTM to represent short-term patterns and integrated by an outer LSTM to model longer-term crop development.*